# Using Electron Energy-Loss Spectroscopy to Measure Nanoscale Electronic and Vibrational Dynamics in a TEM

Ye-Jin Kim, Levi D. Palmer, Wonseok Lee, Nicholas J. Heller, and Scott K. Cushing*

Division of Chemistry and Chemical Engineering, California Institute of Technology, Pasadena, California 91125, United States

*Correspondence and requests for materials should be addressed to S.K.C. (email: scushing@caltech.edu)

**ABSTRACT**

Electron energy-loss spectroscopy (EELS) can measure similar information to X-ray, UV-Vis, and IR spectroscopies but with atomic resolution and increased scattering cross sections. Recent advances in electron monochromators have expanded EELS capabilities from chemical identification to the realms of synchrotron-level core-loss measurements and to low-loss, 10–100 meV excitations such as phonons, excitons, and valence structure. EELS measurements are easily correlated with electron diffraction and atomic-scale real-space imaging in a transmission electron microscope (TEM) to provide detailed local pictures of quasiparticle and bonding states. This perspective provides an overview of existing high-resolution EELS (HR-EELS) capabilities while also motivating the powerful next step in the field – ultrafast EELS in a TEM. Ultrafast EELS aims to combine atomic level, element specific, and correlated temporal measurements to better understand spatially specific excited state phenomena. Ultrafast EELS measurements also add to the abilities of steady-state HR-EELS by being able to image the electromagnetic field and use electrons to excite photon-forbidden and momentum-specific transitions. We discuss the technical challenges ultrafast HR-EELS currently faces, as well as how integration with *in situ* and cryo measurements could expand the technique to new systems of interest, especially molecular and biological samples.

# I. INTRODUCTION

Ultrafast spectroscopy is racing towards the measurement of atomically resolved, element specific, electronic and vibrational dynamics.[1-6] Nanometer resolution optical spectroscopy methods often rely on complex tip-enhanced or nonlinear optical approaches. Similar spectroscopic information can instead be obtained from the inelastic scattering of electrons in a transmission electron microscope (TEM) – an instrument with inherent atomic resolution.[7] High-resolution electron energy-loss spectroscopy (HR-EELS) is enabled by advances in source and monochromator technology.[8-13] HR-EELS effectively turns a TEM into a spectrometer that mimics photon-based measurements, but since electrons possess a larger momentum and a monopolar charge, a wider range of transitions can be accessed including optically-dark, dipole-forbidden modes and momentum non-conserving excitations.[14-16]

Modern day HR-EELS can measure a material's steady-state electronic and chemical properties with meV energetic resolutions. This includes so called low-loss (<50 eV) excitations like phonons or vibrational modes,[9,10] bonding states,[13,17] surface plasmons,[12,18] and the valence bonding or density of states (**Fig. 1(a)**).[19,20] The core-loss region (>50 eV) measures the oxidation state and local coordination of atoms similar to X-ray spectroscopy.[21,22] When HR-EELS is combined with a TEM, it transforms into a more comprehensive spectroscopy, allowing atomic-scale information to be correlated with the element specific, electronic and structural properties of a material. Traditionally, the tradeoff of photon versus electron-based spectroscopy was sample damage, but low-dose and cryogenic approaches now allow measurements of beam-sensitive systems such as molecules, soft polymers, and biological complexes.[23,24]

Extending HR-EELS to the time domain is a powerful research direction. Time-resolved HR-EELS could measure atomic-scale energy transport and local structural dynamics across the range of excitations mentioned. The base technology needed for ultrafast EELS enables correlated ultrafast electron diffraction and real-space imaging techniques in the same ultrafast electron microscope (UEM),[25] the cumulative goal of many large scale X-ray user facilities. Ultrafast EELS also has further distinct advantages such as increased scattering cross sections for measuring light elements or dopants and defects, the ability to directly image the local electromagnetic field for nanophotonics, and the ability to access dipole-forbidden and momentum-specific excitations. That is not to say that true HR-EELS in a UEM is easy to achieve; optimization of photoelectron sources, electron optics, monochromators, and electron detectors are all needed for continued technique growth.

This perspective aims to introduce HR-EELS to a broader audience to motivate the continued development of ultrafast EELS. We briefly introduce the basic principles of EELS in a background section before moving on to surveying the current field of steady-state HR-EELS. We then describe recent developments in the field of ultrafast EELS to show the technique's promise for a variety of materials and nanophotonic systems. Most ultrafast EELS experiments to date have been performed in the low-loss region because of its larger signal intensities and lower demand on spectral resolution.[27-29] We end on a perspective of new frontiers towards time-resolved molecular spectroscopy and *in situ* photochemical dynamics, as well as the technical developments needed to continue to push the boundaries of ultrafast EELS.

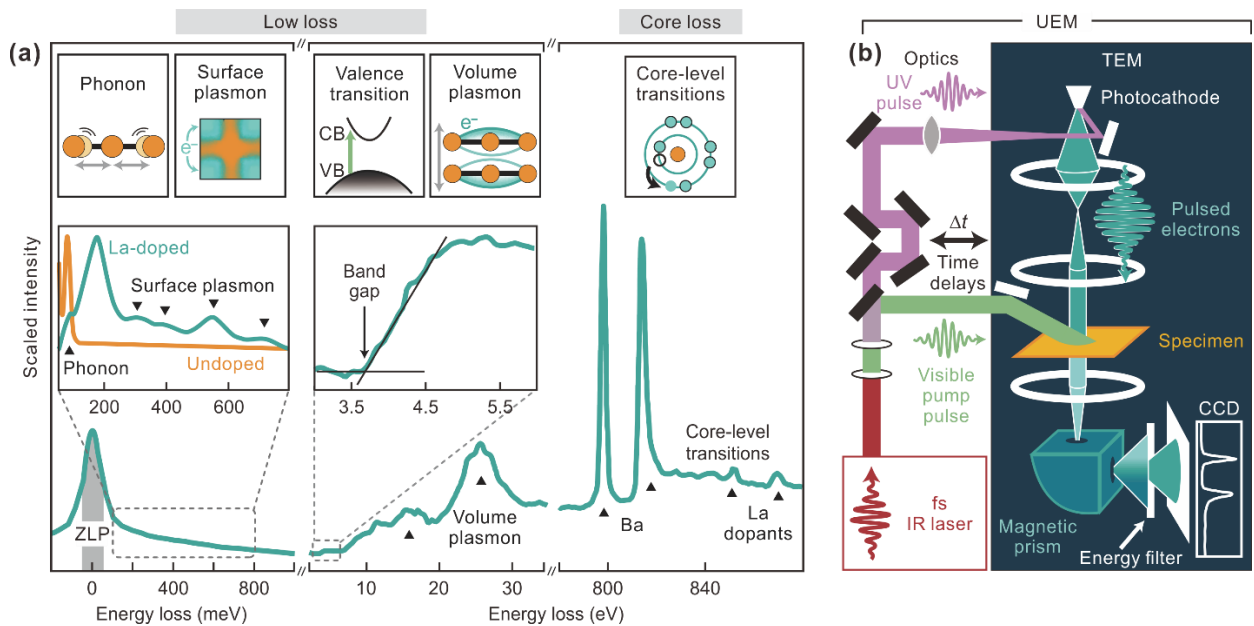

**Figure 1. Ultrafast and high-resolution EELS (HR-EELS). (a)** Steady-state HR-EEL spectra of a 5% La-doped BaSnO$_3$ nanocube (blue spectra) shown in logarithmic vertical scale and the measured excitations. Phonons are visible in the spectrum of the undoped specimen (orange spectra), but they are screened when dopants are introduced. Adapted with the permission from Ref. 26. **(b)** In an ultrafast electron microscope (UEM), femtosecond UV pulses trigger the photoemission from the photocathode, while a time-delayed ($\Delta t$) optical pulse photoexcites the sample. The time-dependent inelastic scattering events of electrons, real-space images, or electron diffraction patterns are then obtained by a charge-coupled device (CCD).

## II. BACKGROUND

### A. Inelastic Electron Interactions in EELS

When free electrons interact with a material, elastic and inelastic scattering occur. Elastic scattering is when electrons are deflected by the Coulombic field of atomic nuclei and no energy is lost. Elastic scattering is primarily associated with diffraction. The diffracted electrons measure momentum-specific structural information due to wave interference according to Bragg's law. Inelastic scattering, as measured with EELS, occurs when the electrons transfer energy and momentum to the sample by excitation of different transitions. Phonons, plasmons, magnons, and inter-/intraband or core-level transitions are examples of possible excitations. Secondary inelastic processes can be measured using energy dispersive X-ray or cathodoluminescence spectroscopies, but they are not discussed in depth here.

EELS is measured by collecting transmitted or reflected electrons and dispersing their energy with a magnetic prism spectrometer (**Fig. 1**). The resultant spectrum contains an elastically scattered zero-loss peak (ZLP) and two regions with different types of excitations, broadly termed low- and core-loss (**Fig. 1(a)**). The full width at half-maximum (FWHM) of the ZLP defines the energy resolution of an EELS measurement. The ZLP typically dominates the EEL spectrum, so the low-loss features are difficult to resolve before the ZLP is subtracted. The signal intensity decreases with increasing energy because the scattering cross section ($\sigma$) is inversely proportional to the energy loss ($E$) of the electrons, as described by the relation, $d\sigma/dE \approx (4\pi a_0^2 R^2/T)(1/E)$, where $a_0$, $R$, and $T$ are the first Bohr radius, the Rydberg energy, and the nonrelativistic incident energy of electrons, respectively.[7]

Within the EEL spectrum, the low-loss region corresponds to vibrational and valence excitations. These excitations often include inter-/intraband transitions and various forms of longitudinal excitations such as surface and volume plasmons, the latter of which are generally inaccessible with photons outside of certain techniques like resonant inelastic X-ray scattering.[30,31] Core-loss EELS provides similar information as X-ray measurements such as atomic bonding configurations,[32] oxidation,[21] and spin states.[33] For instance, core-level spectra indicate configurational changes to a metal oxide's crystal field.[34] EELS is more sensitive to light elements than photons because the scattering cross section of electrons is generally larger than that of photons, which is particularly useful for measuring elements like lithium in batteries or defects, impurities, and dopants. The penetration depth of electrons (usually 1–100 nm) also makes EELS ideal for measuring nanoparticles and two-dimensional or layered materials.

**B. The Difference Between EELS and Photon-Based Spectroscopies**

A simple theoretical model can illustrate the difference between EELS and most photon-based spectroscopies. Starting with Fermi's golden rule, the transition rate $W_{i,f}$ from an initial state $|\Phi_i\rangle$ to a final state $|\Phi_f\rangle$ is given by

$$W_{i,f} = \frac{2\pi}{\hbar} \langle \Phi_f | T | \Phi_i \rangle^2 \delta(E_f - E_i - \Delta E). \tag{1}$$

Here, $\hat{T}$ is the transition operator and the delta function ($\delta$) implies energy conservation. The transition operator for inelastic scattering, can be approximated as[35]

$$\hat{T} = e^{i\mathbf{q}\cdot\mathbf{r}} = 1 + i\mathbf{q}\cdot\mathbf{r} - \frac{1}{2!}(\mathbf{q}\cdot\mathbf{r})^2 + \cdots. \tag{2}$$

For photons, the momentum is small, and therefore **q** and **q** · **r** are small, so terms except the first-order dipole term can be mostly ignored. For electrons, the momentum transfer is significant, and the higher order terms cannot be ignored. For example, measured EEL spectra of a 50 nm SrTiO$_3$ nanoparticle are shown in **Fig. 2**, compared to calculated spectra using a Bethe-Salpeter equation based simulation code (OCEAN).[36,37] When $q$ approaches zero the spectrum resembles X-ray absorption spectroscopy. However, as the momentum transfer increases, the EEL spectrum begins to differ as monopolar and quadrupolar transitions become more significant, as shown by the calculated spectra.[38,39]

It should be noted that **Eq. (2)** assumes the plane-wave Born approximation, valid for electrons transmitted through nanometer-thin specimens. A distorted-wave Born approximation must be used for reflected electrons.[40]

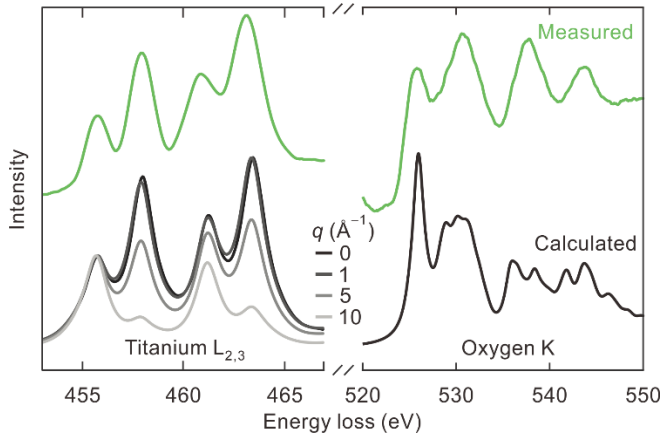

**Figure 2. Experimental and calculated EEL spectra of SrTiO$_3$.** EELS measurements (top) of SrTiO$_3$ nanoparticles are compared to OCEAN calculations (bottom). The edges were acquired at small collection angles at zero momentum ($q = 0$ Å$^{-1}$) to match an X-ray absorption spectrum according to the dipole approximation. Finite momentum transfer in EELS is also calculated as shown in different colors. Momentum is denoted in the panel. The experimental spectrum was smoothed, while the calculated one was broadened to match the experiment.

## III. HR-EELS IN A TEM

HR-EELS is already being used to measure a wide range of excitations with atomic resolution. Here, we discuss a few cutting-edge examples to motivate the on-going adoption of HR-EELS and to motivate its development into an ultrafast technique as discussed in **Sec. IV**.

HR-EELS is approaching measurements of >10 meV excitations with few-meV spectral resolution. Understanding how vibrational modes change at sub-atomic spatial resolutions is therefore a rapidly growing area of spectroscopy. Macroscopic HR-EELS has been used to measure vibrational modes of molecules on surfaces[41] and phonon or plasmon dispersions in momentum space.[40,42-46] The ability to perform HR-EELS in a TEM, however, has only recently become commercially available. High-end TEMs with electron monochromators can now measure EELS with sufficient resolution to differentiate isotope-shifts like in a carboxylate group stretching mode[13] or the differences between bulk and localized vibrational modes around a single substitutional silicon atom in single-layer graphene (**Fig. 3(a)**).[9] Momentum-mapping of vibrational modes, akin to neutron scattering, is also possible with near atomic resolution. Optical and acoustic phonon dispersions in the 50–200 meV range have been mapped in both momentum and real spaces for graphene.[47] Localized, interfacial phonon dispersions from broken translational symmetry were also measured in the atomic junction between boron nitride and diamond.[48] As HR-EELS in a TEM continues to progress, the measurement of localized phonons and their involvement in processes like interfacial thermal transport, superconductivity, topological states, or molecular bonding will continue to be a rapidly growing field.

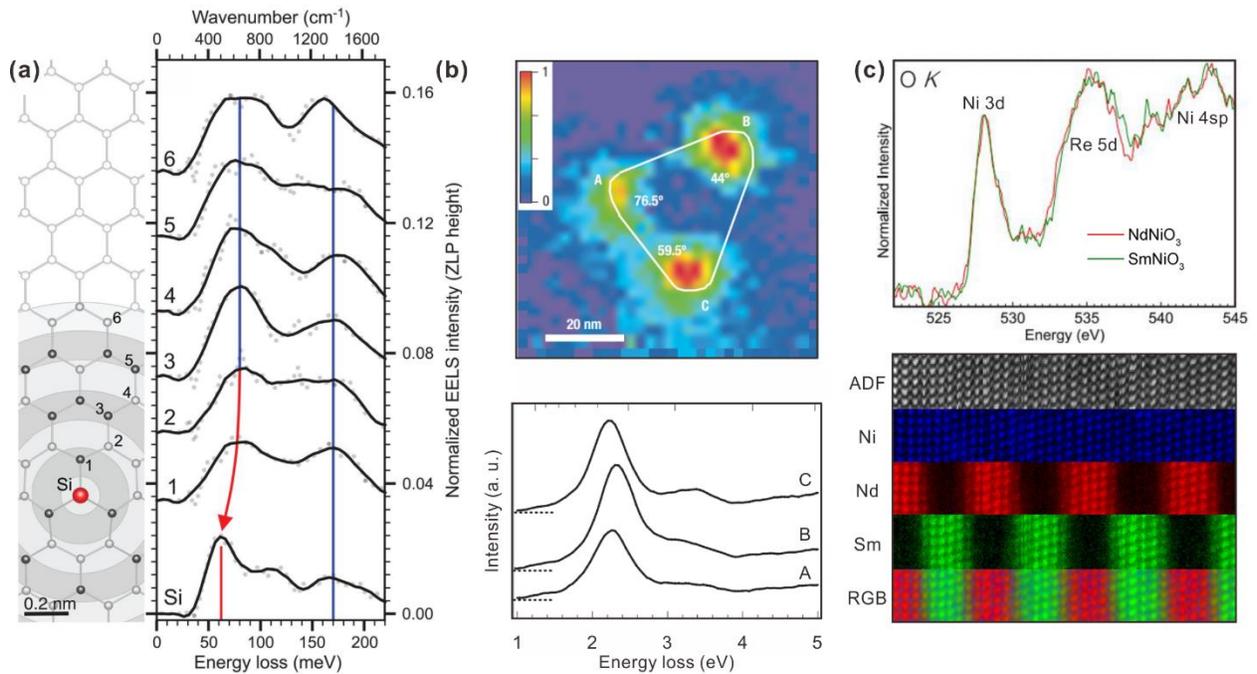

**Figure 3. Steady-state HR-EELS. (a)** Low-loss and vibrational HR-EELS measure localized vibrations around a single silicon impurity in graphene. Each spectrum is measured at the labelled atomic position on the left. Reprinted with the permission from Ref. 9. **(b)** Imaging localized plasmon modes in low-loss HR-EELS. (Upper) EELS probability map of plasmon fields in a silver nanoprism. (Lower) Corresponding EEL spectra for each corner of the nanoprism. Reprinted with the permission from Ref. 12. **(c)** Core-loss imaging of NdNiO$_3$ and SmNiO$_3$ layers. (Upper) Each layer's oxygen K edge spectrum shows the hybridization of the O 2$p$ orbitals with surrounding atoms, Re = Nd or Sm. (Lower) Atomic resolution energy-filtered imaging visualizes the composition and hybridization across the specimen. Reprinted with the permission from Ref. 49.

Low-loss HR-EELS, however, is advantageous for measuring more than phonons. The same energy resolution advances allow for measurements of quasiparticle and valence excitations at the level of optical spectroscopy but now with atomic resolution. For example, the localized nature of exciton-phonon and exciton-plasmon couplings in Moiré superlattices of stacked van der Waals layers has been measured.[50,51] Low-loss HR-EELS has the potential to provide myriad insight into how atomic-level defects like vacancies, interfaces, and local strain modify electronic states and transport. The mass and charge of the electron allow for monopolar and longitudinal excitations to be measured. For example, the photonic density of states and surface or bulk plasmons have been imaged with atomic detail (**Fig. 3(b)**).[12] Low-energy quasiparticles including phonons, polaritons, and excitons, and how they hybridize with photons, have also been measured.[12,18]

In contrast to the low-loss region, the core-loss region relates more closely to X-ray spectroscopies (**Fig. 2**). Electron monochromators and cold field emission guns (cFEGs) in TEMs now match or exceed the spatial and energy resolution of comparable synchrotron techniques. This allows EELS in a TEM to locally differentiate element specific, electronic or structural effects at interfaces,[49] defects,[52] and dopants.[26,53] In **Fig. 3(c)**, near-atomic-resolution, energy-filtered HR-EELS imaging identifies the elemental distribution of few-atom $NdNiO_3/SmNiO_3$ layers. The metallicity of the material and interface is quantified by the energy shift of the Ni $L_{2,3}$ edge.[49] Core-level spectral imaging of La dopant concentrations, down to 1%, was previously demonstrated for a $Ba_{1-x}La_xSnO_3$ nanocube in **Fig. 1(a)**.[26]

## IV. ULTRAFAST EELS

The recent advances in HR-EELS make ultrafast HR-EELS an enticing next step. Ultrafast EELS could probe highly local dynamics and energy transport across the complete vibrational to core loss energy range. Photoinduced dynamics in strain-induced phase separation, charge carrier trapping at atomic defects or doping sites, local state effects in quantum materials, and bond formation in molecular and biological systems are all tempting targets. While technological challenges remain for true ultrafast HR-EELS, see **Sec. V**, rapid progress is being made within this field.

To date, ultrafast EELS has primarily studied surface plasmons and optical near fields because they lead to stronger signals than most low- and core-loss excitations. For example, the ultrafast coherent response from the $\pi + \sigma$ plasmon peak of layered graphite at 20–30 eV is shown in **Fig. 4(a)**.[54] The coupling of the incident photons to induced local charge density appears as quantized equidistant side peaks around the ZLP, usually referred to as photon-induced near-field electron microscopy (PINEM) (**Figs. 4(b)** and **(c)**).[27] PINEM provides a time-resolved picture of the local electromagnetic field, and has led to various applications measuring electron-photon entanglement, quantum optics and photonics,[29,55] and attosecond electron pulse generation.[56] It should also be noted that PINEM is more sensitive to sub-surface plasmonic fields than other photoexcitation based techniques, such as time-resolved photoemission spectroscopy, because electron scattering is measured.[57]

The next logical target for ultrafast EELS is measuring changes in core-loss peaks to mimic ultrafast X-ray measurements but with increased spatial resolution. As technology progresses, these measurements can be correlated with low-loss phonon and valence excitations, but even as is, ultrafast EELS will excel in measuring nanoscale junctions, layered materials, and interfaces[58,59]. Low-dose photoelectron pulses and cryo stages could also extend the technique to molecular and biological level. As of now, ultrafast core-loss EELS is at its first steps, mainly without high spatial selectivity. For example, **Fig. 5(a)** shows a measurement of graphite, where carbon atoms exhibit intralayer $\sigma$-bonds with $sp^2$-hybridized orbitals and interlayer $\pi$-bonds with $p_z$ orbitals, each with different energy-loss peaks.[60] **Figures 5(b)** and **(c)** then show the ultrafast EEL spectra of graphite's carbon K edge upon photoexcitation. The transient feature around 290 eV corresponds to a red-shift of the $\sigma^*$ peak and indicates an elongation of the $\sigma$-bonds in the basal plane. The charge-transfer dynamics in hematite ($\alpha$-$Fe_2O_3$) nanoparticles have also been measured. Photoexcited $Fe^{2+}$–$Fe^{4+}$ electron-holes generated from Fe $3d$ transitions between nearby atoms were used to spatially map the excitations within a nanoparticle at several time points.[61]

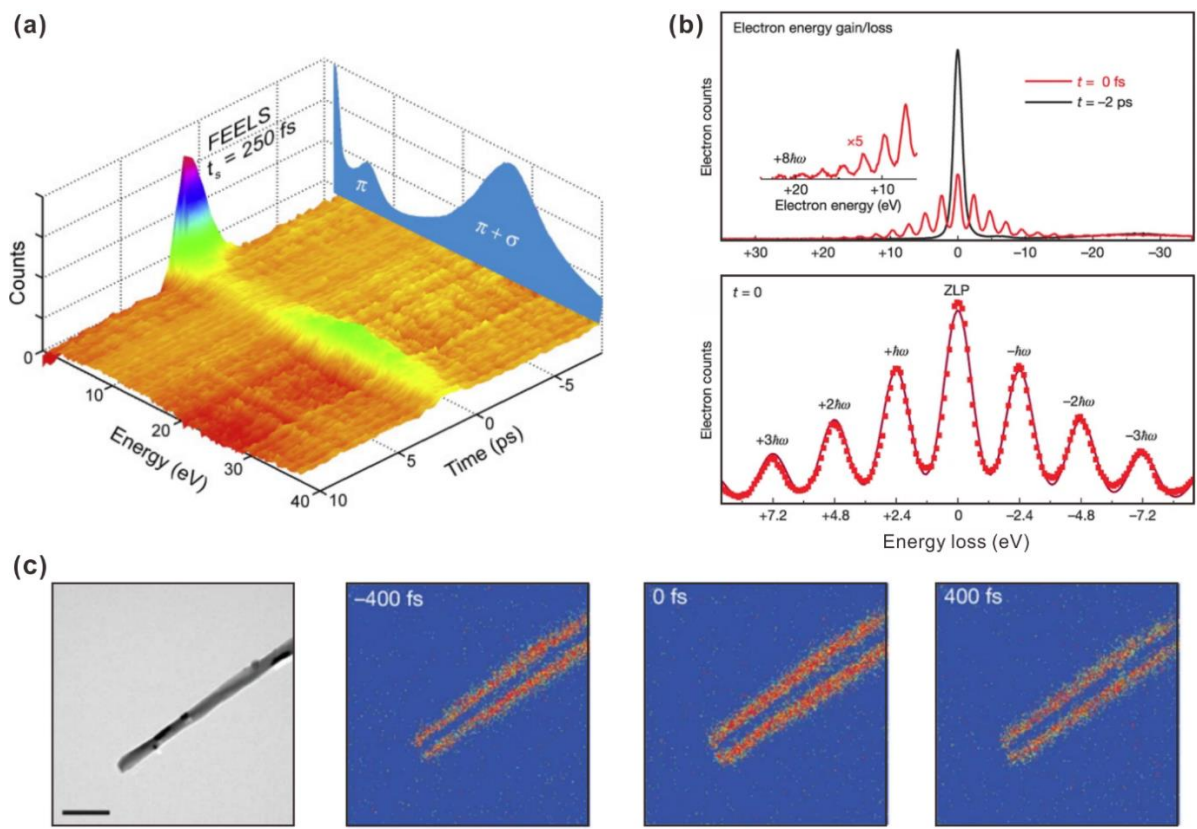

**Figure 4. Ultrafast low-loss EELS in UEM. (a)** Ultrafast EEL spectra of a photoexcited graphite thin film. Time-dependent modulation of $\pi$ and $\pi + \sigma$ plasmon peaks is correlated with interlayer shear motions. Reprinted with the permission from Ref. 54. **(b)** Photon-induced near-field electron microscopy spectra and **(c)** corresponding energy-filtered images of carbon nanotubes. Near-field electron-photon interactions manifest as the gain and loss side bands. Scale bar, 500 nm. Adapted with the permission from Ref. 27.

The inherent combination of ultrafast EELS and time-resolved electron diffraction in UEM is another important motivator, as it allows for easier separation of electronic and structural effects in the core-loss spectrum. For example, a correlated approach was used to determine the photoinduced structural and electronic phase changes in bi-layered manganite ($PrSr_{0.2}Ca_{1.8}Mn_2O_7$) (**Figs. 5(d−f)**).[62] Comparing ultrafast low- and core-loss EELS with crystallographic information confirmed transient changes to the lattice in three different crystalline axes (**Fig. 5(d)**). This combined information assigned that a photothermal pressure wave was inducing coherent oscillations in the manganite lattices (**Figs. 5(e)** and **(f)**). **Figure 5(f)** also shows that the Mn $M_{2,3}$ edges are the most sensitive to the photothermal pressure wave based on their larger oscillation amplitudes than the others, hinting at the electronic structure.

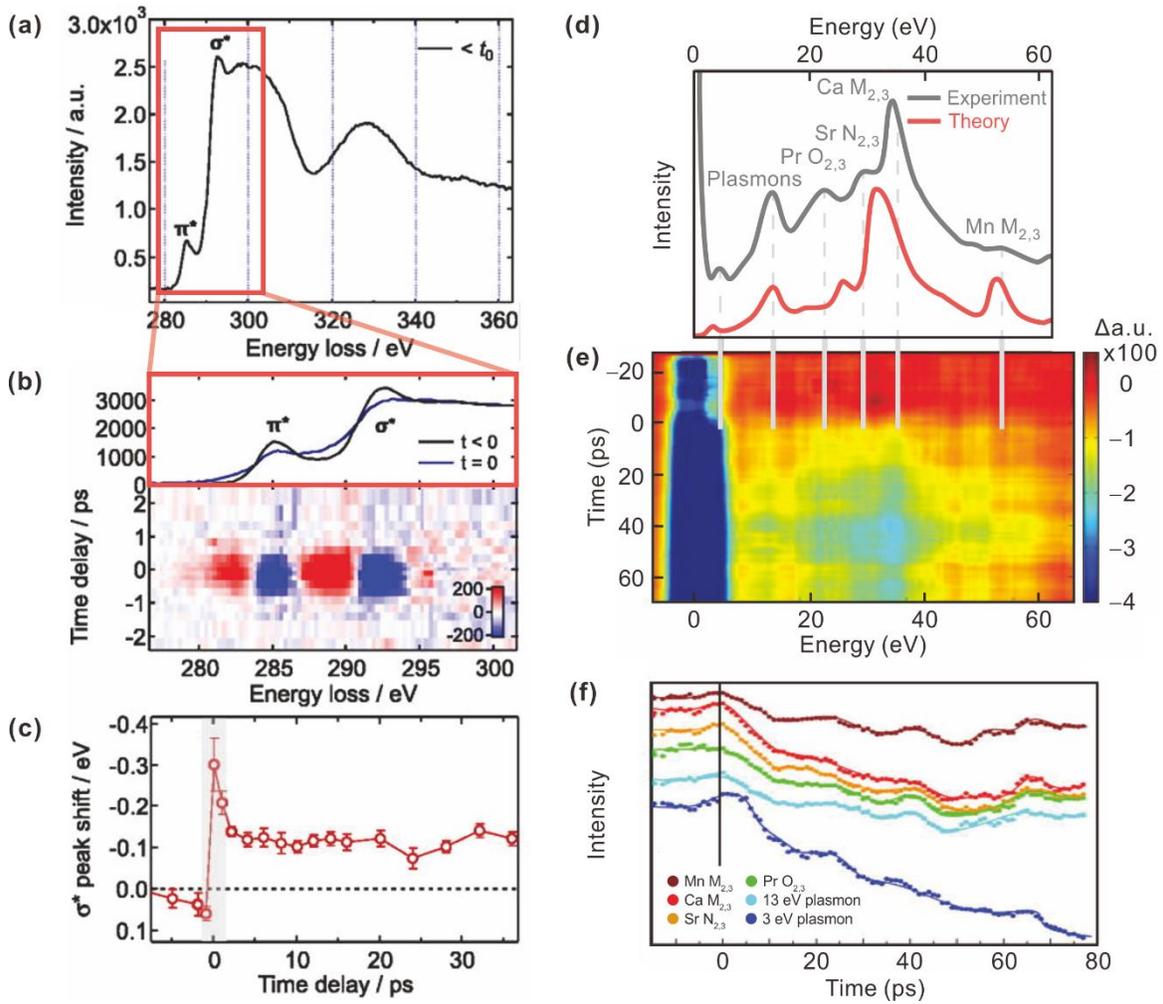

**Figure 5. Ultrafast core-loss EELS in UEM. (a)** Static core-loss EELS of graphite. Two major spectral features correspond to transitions from the $1s$ core-level to anti-bonding $\pi^*$ and $\sigma^*$ states. **(b)** Ultrafast core-loss EELS of graphite. Electron energy loss/gain during interaction with a photon pulse broadens the spectrum around $\Delta t = 0$ (top) and obscures dynamics in the first picosecond (bottom). **(c)** The $\sigma^*$ peak position temporal evolution. The gray shaded area indicates the PINEM lifetime. Adapted with the permission from Ref. 60. **(d)** Static experimental (grey) and simulated (red) EEL spectra of a $PrSr_{0.2}Ca_{1.8}Mn_2O_7$ film. The grey dashed lines guide the eye to the peak positions in the transient ultrafast map. **(e)** A differential map of photoexcited bi-layered manganite, including low energy core levels and plasmon dynamics. The plot is obtained by taking the difference between $EELS(t) - EELS(t<0)$. **(f)** Time-dependent EELS intensity profile for the plasmon and core-level excitations as labeled in panel **(d)**. Adapted with the permission from Ref. 62.

# V. TECHNOLOGICAL HURDLES FACING ULTRAFAST EELS

The next section describes, for the more expert reader, the technology development needed for ultrafast EELS. A non-expert reader can feel free to skip ahead to the next section on new frontiers for the field without losing any context.

The long-term potential of ultrafast HR-EELS to measure chemical, electronic, and vibrational dynamics with nanometer spatial resolution, alongside complementary electron diffraction and real-space imaging, makes ultrafast EELS in UEM a high priority for continued scientific development. However, technological advances are still needed to make the described ultrafast HR-EELS measurements more feasible, especially through shorter acquisition times, but also to reach the spectral and spatial resolution of steady-state variants. For reference, **Fig. 6(a)** gives an estimate of the energy ranges needed to measure different phenomena in ultrafast TEM. Ultimately, the performance metrics of ultrafast EELS, in or out of a TEM, is set by the photocathode source, the electron monochromatization, and detector sensitivity. For the source, high current and coherence of an electron beam with low transverse emission angle, would be demanded. A high beam current in a small spot size leads to realistic acquisition times versus system drift for spatiotemporal measurements. Ideally, a high-current photoelectron source would be integrated with an electron monochromator, in addition to aberration-corrected electron optics, but these often create conflicting design criteria. Post-specimen detection, such as an aberration-corrected EEL spectrometer and a direct electron detector with a high quantum efficiency and large dynamic range, is slightly easier to achieve. The realistically obtainable spectral resolution in ultrafast EELS will be a compromise of these criteria. The following sections break down the major technological advances needed to progress towards ultrafast EELS in more detail.

## A. Photoelectron Sources

It is widely acknowledged that the photoelectron source is currently a major limiting factor for the performance of UEM. Electron emission mechanisms of TEM cathodes fall into two broad classes: thermionic and field emissions (**Fig. 6(b)**). Thermionic emission produces a broadband, thermal electron distribution above the cathode's work function with a 0.7–1.5 eV FWHM bandwidth. Schottky FEGs decrease the spectral width to 0.5–1.1 eV by applying an electric field that extracts electrons from a nanoscale tungsten tip coated with zirconium oxide (ZrO/W), but they are still slightly heated. The newest tip technology are cFEGs, which, as the name implies, use a non-heated FEG tip. These nanostructured cathodes are kept at ambient temperature in a strong electric field and result in the lowest emission bandwidths (<0.4 eV) and highest coherence.[63,64] When cooled to <10 K, even <15 meV bandwidths are possible.[65] An electron monochromator combined with a cFEG cathode enables meV energy resolution.[66,67] Microwave cavity can also be a viable alternative to photocathodes to generate pulsed electrons.[68,69] However, electronic and thermal relaxation may not fully occur between laser pulses of MHz frequency, leading to the buildup of a non-equilibrium background, and changing the measured dynamics.

Ultrafast EELS typically relies on photoemission from a cathode instead of applying thermal heating or electric fields. Laser excitation places different constraints on the tip design that affect the energy resolution and the EELS performance. The profile of the incident light on the photocathode influences the properties of the photoelectron pulse. There are typically two types of photocathodes (**Fig. 6(c)**). Flat tips are the most widely employed because they produce a higher photoelectron current. Common tip materials include lanthanum hexaboride, tantalum, and thin films of gold. Photocathodes with radii ranging from tens to hundreds of nanometers have a lowered work function, like Schottky FEGs, and an increased extraction efficiency due to the local electric field. Nanotips also have a reduced electron emission cone angle.

For example, a resolution of approximately 200 fs and 0.6 eV was achieved using a side-illuminated ZrO/W nanotip with a radius of 10–100 nm at a focused beam size of 9 Å.[70] Despite the significant combined resolutions in this example, using nanoscale photocathodes remains challenging in most UEMs. This is largely due to the lower photoelectron flux produced upon illumination – on the order of ten to a hundred times fewer electrons – compared to micron-sized cathodes. In PINEM, where strong pump-probe interactions are present, the use of a nanotip can enable a more coherent photoelectron beam and facilitate stronger interactions with the incident optical pulse. When acquiring core-loss spectra, a higher electron flux is desirable because of the smaller scattering cross section, so the balance of nano- versus macro- sized tips is again different. Monochromators, as discussed in the next, may therefore be the key component.

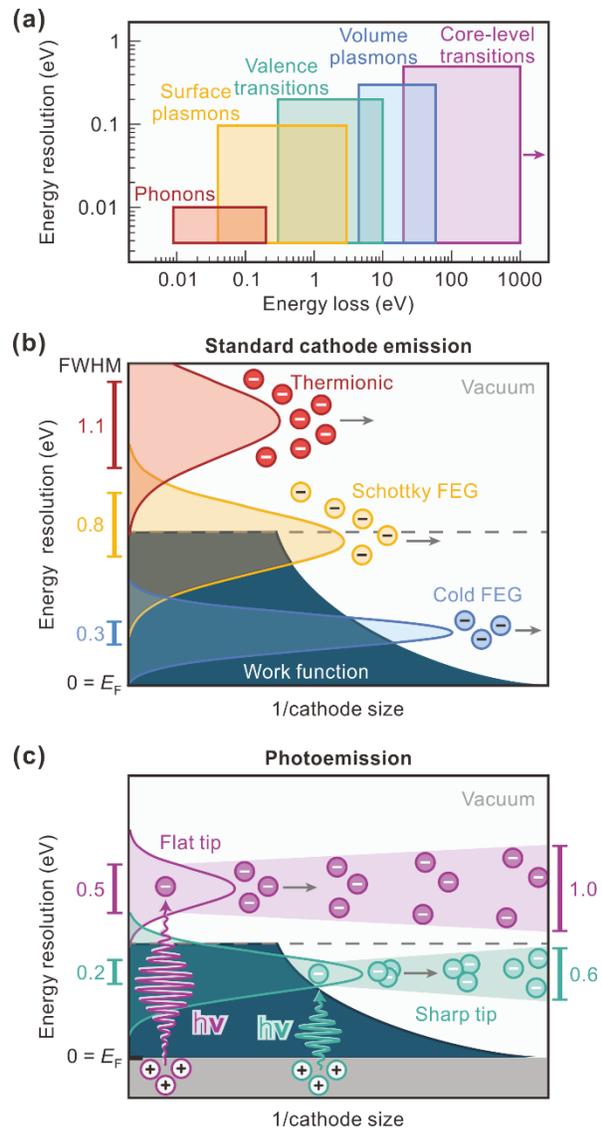

**Figure 6. Cathode types and EELS resolutions. (a)** Energy resolution of an electron beam versus inverse cathode size. Thermionic gun or field emission gun, FEG, generate an electron distribution in vacuum by extracting electrons around the Fermi level ($E_F$). The average emission bandwidth (FWHM) of each cathode is determined by its work function, which decreases as the size of the cathode decreases. The dashed line depicts the bulk work function. **(b)** Photocathodes (flat or sharp tips) are photoexcited with a femtosecond laser. The photoelectron pulse broadens during propagation. **(c)** The energy loss region and energy resolution required to elucidate dynamics with EELS. Each excitation's fine structure peak separation and spectral bandwidth determine the resolution required.

**B. Electron Monochromators**

While electron monochromators have been employed for several decades, their integration into aberration-corrected and high-acceleration-voltage TEMs represents a recent advancement. Monochromators can now reduce the electron beam energy width down to <10 meV.[67] However, increasing spectral resolution comes at the cost of beam brightness. Standard monochromators, such as a Wien filter or electrostatic omega filter, usually decrease beam brightness by two orders of magnitude and introduce additional beam blur due to stochastic electron-electron interactions. Photocathode currents are already several orders of magnitude lower than continuous-wave electron beams, so this presents a challenge for ultrafast EELS.

Recent advances attempt to increase spectral resolution without compromising the photoelectron flux. Radiofrequency (RF) accelerating cavities improve the energy resolution of pulsed electrons without loss by re-aligning the electrons in longitudinal phase space. Theoretical studies suggest that time-correlated acceleration in a pair of RF cavities can equalize the energy distribution of the dispersed electron beam to achieve monochromatization.[71] Tunable electromagnetic MeV monochromators with bending magnets have demonstrated $10^{-5}$ energy selection while maintaining a sub-nanometer spatial resolution in a UEM.[72] This resulted in a monochromator efficiency of 13% for a few picosecond photoelectron pulse. It is also possible to design a lossless monochromator by interacting the photoelectrons with an external single-cycle THz pulse (**Fig. 7(a)**).[73-75] Integrating a THz compressor or laser-driven THz emitter along the electron beam path can decelerate high-energy leading electrons and accelerate low-energy tail electrons, resulting in a narrower energetic and temporal distribution, but at the added complexity of synchronizing with the additional external field. Significant additional research is still needed for these directions before mass integration with ultrafast electron microscopes will be possible, making it an important area of continued effort.

**C. Electron Lenses and Detection Systems**

The performance of ultrafast EELS can also be improved by optimizing the electromagnetic lens system to reduce aberrations and mitigate photoelectron pulse broadening at beam crossover. The photoelectron source and the electrostatic potential of the photocathode and photoanode are often considered the most deterministic factors for spatiotemporal and energy resolution.[76-78] For example, spherical aberrations are negligible for the pulsed electrons in objective lenses.[79] The stack of multiple electromagnetic lenses at high magnifications, however, has been found to create temporal distortions because of the varying trajectories of photoelectrons at the focal plane.[80]

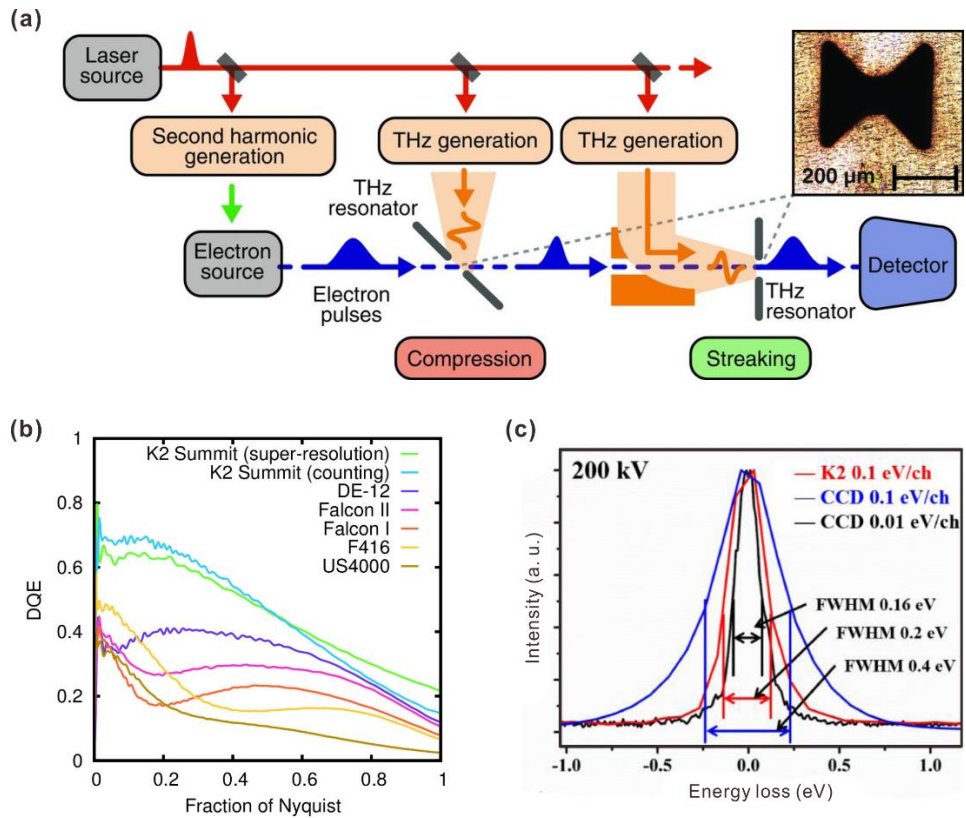

**Figure 7. Monochromatization and sensitivity improvements for ultrafast EELS. (a)** Photoelectron pulse compression with few-cycle THz pulses. Phase and shape of the ultrafast electron pulses are shaped and controlled using a THz field by decelerating and accelerating leading and trailing electrons, respectively. The butterfly-shaped metal resonator mediates the interaction between the electrons and THz fields. Both the temporal and energy bandwidths of the photoelectron pulse are subsequently narrowed. Reprinted with the permission from Ref. 75. **(b)** Detection quantum efficiency (DQE) of different detector technologies. Direct electron detection devices such as the K2 Summit detector and DE-12 have higher quantum yield than traditional CCD cameras. Reprinted with the permission from Ref. 81. **(c)** A ZLP comparison between a direct electron detection camera and a CCD. For the same spectrometer dispersion and acquisition setting, direct electron detection leads to a narrower ZLP, and thus a better spectral resolution. Adapted with the permission from Ref. 82.

    The spectral resolution and sensitivity of ultrafast EELS are most easily improved by incorporating more efficient electron detectors. High quantum efficiency and dynamic range are needed so that small differential signals can be distinguished from ground-state spectral features. Improved detectors are particularly important in core-loss spectroscopy at high energies where the scattering cross section is smaller. In recent years, direct electron detection cameras have significantly improved the spectral sensitivity and frame rates over scintillator-based technology.[83-85] These detectors also enable low-dose imaging of organic and biological materials.[86,87] In conventional scintillator-coupled CCDs, electron and photon scattering produce

delocalized signals which increase the point spread function of the detector, which in turn, reduces spectral and spatial resolution. Direct electron detection cameras simultaneously readout electron exposure, improving the point spread function as well as the processing speed and quantum efficiency.[81,84] Several studies have quantified the advantage of direct electron detection in EELS (**Figs. 7(b)** and **(c)**).[82,88] High-resolution, time-resolved imaging in a single-electron-per pulse regime has also been demonstrated.[89] Advanced electron sensors therefore already offer significant benefits for ultrafast EELS in existing systems, although still costly to implement.

## VI. NEW FRONTIERS: MOLECULES AND *IN SITU* CELLS

To date, most EELS applications focus on thin films or nanoparticles. However, integration with cryo, *in situ*, and low-dose methods now promises measurements of molecular and bio-molecular systems. Since HR-EELS can resolve vibrational modes of chemical bonds analogous to IR spectroscopy (**Fig. 8(a)**),[24,90] and also label chemical isotopes,[13] the nanoscale spatial resolution may be able to differentiate processes like local protein-water bond networks or a polymer cross-linker's coordination. Combined with techniques like microcrystal electron diffraction (microED), which determines crystal structures of organic molecules at Å-scale resolution,[91] electronic and chemical structures could be obtained even the mixed molecular products of a reaction.

Similarly, ultrafast HR-EELS will excel in measuring local, bond-specific changes in photoinduced chemical processes. The combined electronic and structural information could be used to understand light-induced conformational changes of proteins or macromolecules in combination with chemical reactions. The technique could ultimately offer molecular-scale insights into chemical kinetics, while investigating real-space distributions that vary with the local micro-environment. From a more technical perspective, ultrafast EELS is also advantageous for investigating molecular systems due to its use of a lower probe dose, which avoids severe knock-on displacement or radiolysis ionization.[92] Direct electron detectors for low-dose measurements,[85] cryogenic environments,[93,94] and encapsulating molecules with a graphene liquid cell are all routes that need further research if EELS is to reach beyond inorganic materials.[86]

We view combining ultrafast and HR-EELS with *in situ* and *operando* measurements as another important frontier. Whether photocatalysis, ion diffusion, or gas phase reactions, EELS has the potential to advance our understanding of myriad systems. Environmental cells can now be used to apply heat or electrical bias in liquid, solid, and gas phases, for which ultrafast HR-EELS could measure the induced chemical product evolution or nanoscale transport in their intrinsic timescales.[95] For example, EELS was used to chemically map the local composition and distribution of molecules in the liquid phase, measuring the local, heterogeneous photoinduced reaction in catalytic nanoparticles (**Fig. 8(b)**).[96] Photocatalysis and photoelectrochemistry are particular areas of interest where HR-EELS and ultrafast EELS could push the boundaries of current knowledge. Most photocatalysts depend on nanoscale effects like facet-specific chemistry, local morphology changes, grain boundaries, dopant distributions, defects, and atomic-scale co-catalysts which lead to contested mechanistic pictures. Photomodulated, steady-state HR-EELS would even give new insight into local reactions and bottlenecks for carrier transport. Eventually, ultrafast EELS could achieve the holy grail of measuring photocatalysis all the way from photoexcitation to transport, to site-specific product formation in terms of local morphology, vibrational modes, and electronic structure.

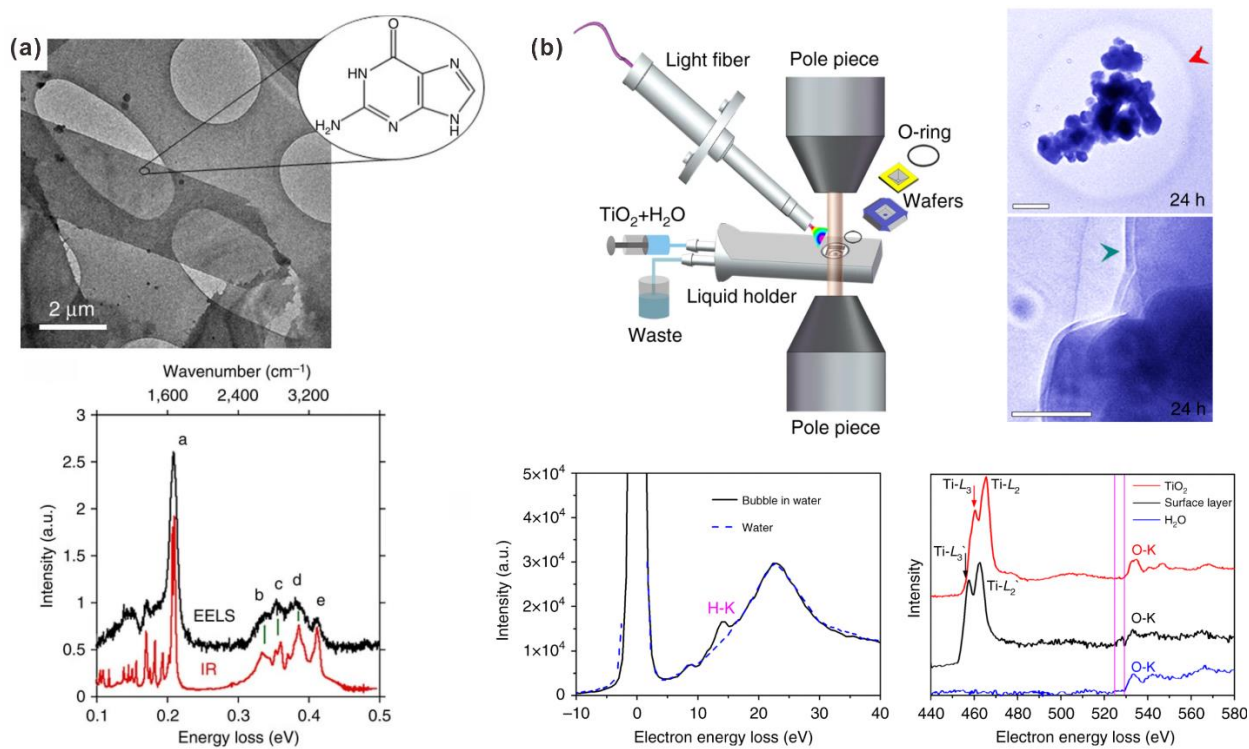

**Figure 8. EELS for molecular and *in situ* systems. (a)** Identification of molecular vibrations with HR-EELS. Vibrational signatures of C–H, N–H, and C=O in a guanine molecular crystal reveal that HR-EELS can be an analogue technique to IR spectroscopy even with the high structural resolution. Adapted with the permission from Ref. 90. **(b)** *In situ* observation of photocatalytic water splitting in $TiO_2$ nanoparticles. Using a fluidic TEM holder under continuous-wave UV light illumination, a hydrogenated shell formed around the nanoparticles. The photocatalytic processes were analyzed with HR-EELS, confirming the shell consists of hydrogen nanobubbles. Valence states of $TiO_2$ are also observed to be reduced (relative intensity ratio of Ti $L_2$ and $L_3$ edges) and the appearance of a pre-edge peak in oxygen K edge around the particle surface is attributed to an unpaired oxygen originating from the loss of hydrogen. Adapted with the permission from Ref. 96.

## VII. SUMMARY AND FUTURE OUTLOOK

Steady-state and photoexcited HR-EELS has the potential to push spectroscopy's limits in energy, space, and time. HR-EELS is slowly becoming available in commercial TEMs. Combining HR-EELS with spatial imaging and electron diffraction in a TEM can provide near complete properties of interfaces, defects, dopants, and nanoscale features. Ultrafast EELS still needs further development because photoelectron pulse broadening restricts the spatial and energy resolutions. High coherence and high flux electron pulses are needed but place demand on advances in photoelectron sources and electron optics. Eventually, energy-filtered imaging with ultrafast EELS will directly visualize the spatiotemporal evolution of photodynamics for excitations like phonons, plasmons, or polarons. We also envision investigating ultrafast chemical reaction dynamics in molecular and photoelectrochemical systems through low-dose methods. Electrical excitation alongside optical excitation pulses would further expand the set of measurable phenomena. Either way, although significant technical developments are needed, ultrafast HR-EELS has the potential to be a defining spectroscopic tool. While combinations of optical and X-ray spectroscopies can access parts of ultrafast HR-EELS measurements, combining all aspects of UEM in one instrument with atomic scale resolution makes the needed technological development worthwhile, and will prove transformative for sciences spanning the energy, materials, and quantum spaces including new applications in the molecular and bio-molecular world.

## ACKNOWLEDGEMENTS

This research was supported as part of the Ensembles of Photosynthetic Nanoreactors, an Energy Frontiers Research Center funded by the U.S. Department of Energy, Office of Science under Award Number DE-SC0023431. Y.-J.K. is further supported by the Liquid Sunlight Alliance (the U.S. Department of Energy, Office of Science, Office of Basic Energy Sciences, Fuels from Sunlight Hub under Award Number DE-SC0021266). L.D.P is supported by the National Science Foundation Graduate Research Fellowship under Grant Number DGE-1745301. W.L. acknowledges further support from the Korea Foundation for Advanced Studies.

## AUTHOR DECLARATIONS

**Conflict of Interest**

The authors have no conflicts to disclose.

## DATA AVAILABILITY

The data that support the findings of this study are available from the corresponding author upon reasonable request.